\documentclass[prb,twocolumn,superscriptaddress,showpacs]{revtex4}

\input{epsf}
\begin{document}

\title{Oscillatory Size-Dependence of the Surface Plasmon Linewidth 
       in Metallic Nanoparticles}
\author{Rafael A.\ Molina} 
\affiliation{Departamento de F\'{\i}sica At\'omica, Molecular y Nuclear, 
            Universidad Complutense de Madrid,  
            28040 Madrid, Spain}
\affiliation{Institut de Physique et Chimie des Mat\'eriaux de Strasbourg,
       UMR 7504 (CNRS-ULP), 23 rue du Loess, 67037 Strasbourg, France}
\author{Dietmar Weinmann}
\affiliation{Institut de Physique et Chimie des Mat\'eriaux de Strasbourg,
       UMR 7504 (CNRS-ULP), 23 rue du Loess, 67037 Strasbourg, France}
\author{Rodolfo A.\ Jalabert}
\affiliation{Institut de Physique et Chimie des Mat\'eriaux de Strasbourg,
       UMR 7504 (CNRS-ULP), 23 rue du Loess, 67037 Strasbourg, France}

%
% PACS
%
% 71.45.Gm is Exchange, correlation, dielectric and magnetic response
%             functions, plasmons.
% 36.40.Vz is Optical properties of atomic and molecular clusters.
% 31.15.Gy is Semiclassical methods in atomic and molecular physics.
     
\begin{abstract}
We study the linewidth of the surface plasmon resonance in the optical 
absorption spectrum of metallic nanoparticles, when the decay into 
electron-hole pairs is the dominant channel. Within a semiclassical 
approach, we find that the electron-hole density-density 
correlation oscillates as a function of the size of the particles, leading to 
oscillations of the linewidth. This result is confirmed numerically for 
alkali and noble metal particles. While the linewidth can increase strongly, 
the oscillations persist when the particles are embedded in a matrix.  
\end{abstract}

\maketitle

% 
% ******************* INTRODUCTION ************************
% 
\section{Introduction}
The lifetime of collective excitations in finite quantum systems is a 
fundamental problem spanning from the physics of nuclei to that of
metallic clusters \cite{bertsch_broglia}. Giant resonances, in 
the former case \cite{Heyde}, and surface plasmon, in the latter
\cite{mie}, decay by different mechanisms according to the physical 
regime of the excitation.
For {\it large}\/ metallic clusters (with radius $R \gtrsim 10\,{\rm nm}$) 
radiation damping is the main limiting factor to the plasmon lifetime 
\cite{vollmer_kreibig}. At {\it small}\/ sizes 
($0.5\,{\rm nm} \lesssim R \lesssim 2\,{\rm nm}$) the decay into
electron-hole pairs (Landau damping) dominates, while for 
{\it intermediate}\/ sizes both effects compete. For {\it very small}\/ 
clusters ($R \lesssim 0.5\,{\rm nm}$) a strong temperature dependence 
of the linewidth \cite{ellert} suggests that the coupling of the 
collective excitation to thermal vibrations of 
the atomic lattice is an important effect.

Except for the case of very small clusters, where the atomic arrangements
play a crucial role, the theoretical analysis of the surface plasmon is 
usually based on the jellium model, where conduction electrons move in a 
uniformly charged background. Within such a framework, the Landau damping 
occurs when there are particle-hole excitations with an energy close to 
that of the collective excitation. Therefore, this mechanism is strongly 
dependent on the density of states of the conduction electrons confined in 
the cluster.

Using a linear response approach, Kawabata and Kubo \cite{kawabata_kubo} 
proposed a total linewidth $\Gamma = \Gamma_{\text{i}} + \Delta \Gamma(R)$, where 
the intrinsic width $\Gamma_{\text{i}}$ is given by bulk properties and the 
size-dependent contribution $\Delta \Gamma(R)$ is proportional to $1/R$. 
Such an $R$ dependence would also result from the identification of the 
diameter of the particle with the electron mean-free-path. However, as 
stressed in the original work, this identification is not completely 
correct, since the quantized states exist due to the confinement and 
cannot be considered to be scattered at the surface 
\cite{kawabata_kubo,bertsch_broglia}. On the other hand, 
for interacting electrons confined in a parabolic potential, Kohn's 
theorem \cite{kohn} implies an infinitely-lived collective excitation 
\cite{dobson}, such that the linewidth of the plasmon can 
be thought as a measure of the failure in separating the collective 
(center-of-mass) and relative coordinates for the electron system.

Further improvements in the linear response theory by 
Yannouleas and Broglia \cite{yann_broglia} and by
Barma and Subrahmanyam \cite{barma}
recovered the $1/R$ dependence
of $\Delta \Gamma(R)$ in the asymptotic limit of large $R$. Numerical 
calculations \cite{barma,yannouleas} and existing experiments 
\cite{vollmer_kreibig,Kreibig} on free in\-ter\-medi\-ate-size clusters are in 
good agreement with the values of $\Delta \Gamma(R)$ yielded by those 
calculations. For small clusters the emergence of a non-monotonous 
size-de\-pend\-ent structure in the lineshape 
(in experiments \cite{charle,brechignac,mochizuki} as well as in 
numerical calculations 
\cite{barma,yannouleas,kleinig,babst,ekardt?}) 
has been attributed to shell effects (see also Fig.~10 of 
Ref.~\cite{yannouleas}a). We provide in this work a
detailed theoretical understanding of these effects and determine 
analytically the size-dependence of the plasmon lifetime for free and 
embedded clusters.

Using a semiclassical approach, we explain how a subtle consequence of 
electronic shell effects, namely angular-momentum dependent 
electron-hole density-den\-sity correlations, leads to an 
{\it oscillatory size-depen\-dence of the plasmon linewidth due to Landau 
damping}\/ in small alkaline metal clusters. This result is 
confirmed by numerical calculations using the time-dependent local
density approximation (TDLDA) \cite{ekardt,yannouleas}. We show that our
results are relevant also for noble metals and clusters embedded in matrices. 
Nanoparticles embedded in a 
matrix are interesting from the experimental point of view since the 
position of the resonance can be tuned by changing the dielectric surrounding. 
They constitute a complicated system, where the plasmon lifetime strongly 
depends on the properties of the matrix, an effect that has been called 
chemical interface damping \cite{Kreibig,Hovel}. The application of the
TDLDA, within a simple model of the potential at the surface, confirms 
that the size-oscillation of the level-width predicted by our 
analytical method persists even for embedded clusters.
%
%%%%%%%%%%%%%%%%% PROBLEM and ANALYTICS %%%%%%%%%%%%%%%%%%%%%%%%%%%%%%
%

\section{Landau damping in nanoparticles}

The nanoparticles we consider are much smaller than the wavelength of the
radiation, and the optical absorption cross-section at frequency $\omega$ 
is given by 
\begin{equation}\label{cross-section}
\sigma(\omega)=\frac{4\pi e^2 \omega}{3c}
\sum_j \left|\langle j|z|0\rangle \right|^2 
\delta(E_j-\hbar\omega) \, ,
\end{equation}
where the dipole matrix element is taken between the electronic
many-body ground state $|0\rangle$ 
and the excited states $|j \rangle$ with energy $E_j$ (the $z$-axis is 
chosen parallel to the electric field direction). The spectrum is dominated by 
the collective surface plasmon excitation which is characterized by an 
oscillation of the center of mass of the electrons with respect to the lattice 
of positive ions. Following Mie's classical theory \cite{mie}, for a
nanoparticle in vacuum one expects the surface plasmon resonance at
the frequency $\omega_{\text{M}}=\omega_{\text{p}}/\sqrt{3}$, where 
$\omega_{\text{p}}$ is the bulk plasmon frequency of the material.

When the confinement is not parabolic the surface
plasmon is not an eigenstate of the many-body Hamiltonian and its
coupling to the particle-hole excitations leads to a broadening of the 
corresponding resonance in the spectrum (see the inset of Fig.~\ref{fig:na}). 

A common approach to the plasmon lifetime consists in treating the collective
excitation as an external perturbation which can give rise to the creation 
of electron-hole excitations \cite{bertsch_broglia}. Within this framework, 
Fermi's Golden Rule yields the linewidth
\begin{equation}
\Delta \Gamma (R)= 2\pi \sum_{ph} \left|\langle p | \delta V | h \rangle 
\right|^2 \, \delta(\hbar\omega_{\text{M}} - \epsilon_p + \epsilon_h)\, ,
\end{equation}
where $| p \rangle$ and $| h \rangle$ are electron and hole states in the 
self-consistent field, respectively, and $\delta V$ is the dipole field due 
to the surface plasmon. More rigorous approaches, like discrete-matrix
RPA would lead after certain approximations \cite{yann_broglia}, to an
equivalent separation of collective and particle-hole states.

In the case of spherical symmetry we can work in the effective one-dimensional 
problem of the radial coordinate for each value (in units of $\hbar$) of the
angular momentum $L< \rho =R\sqrt{2ME}/\hbar$, $E$ and $M$ being the
energy and the effective mass of an electron, respectively, and $R$
denoting the radius of the nanoparticle. Assuming that the confinement and the 
interactions lead to hard walls at radius $R$ in the self-consistent
field, and integrating over the electron-hole states, one obtains 
\cite{yann_broglia}
\begin{widetext}
\begin{equation}\label{gamma_density}
\Delta \Gamma (R) = C \gamma \!\!\! \int\limits_{E_{\text{F}}}^{E_{\text{F}}+
   \hbar\omega_{\text{M}}} \!\!\!{\text{d}}E\, 
  \sum_L \sum_{L'=L\pm 1} (2L+1)(2L'+1)  
<L,0;1,0|L',0>^2 
E (E-\hbar\omega_{\text{M}})d_L(E)d_{L'}(E-\hbar\omega_{\text{M}})\, ,  
\end{equation}
\end{widetext}
where $<L,0;1,0|L',0>$ is a Clebsch-Gordan coefficient, 
$\gamma=(2\pi\hbar^3)/(3NM^2\omega_{\text{M}}R^4)$, $C=4MR^2/\hbar^2$, 
and $d_L(E)$ is the one-dimensional density of states. 

\section{Semiclassical theory}

Our approach consists in the evaluation of (\ref{gamma_density}) by
using the semiclassical expression \cite{gutz_book}
$$
d_L(E)=\frac{\tau(E,L)}{2\pi\hbar}\left(1+2\sum_{r=1}^\infty 
       \cos\left[r\left(\frac{S(E,L)}{\hbar} -\pi\right)\right]\right)\, , 
$$
which decomposes $d_L(E)$ in its smooth and oscillating components. 
Here, $\tau(E,L)=\hbar\sqrt{\rho^2-L^2}/E$ is the period of the 
one-dimensional motion, 
$$
S(E,L)=2\hbar\left[\sqrt{\rho^2- L^2} -L \arccos(L/\rho)\right]
$$
is the corresponding action, and the sum runs over all numbers of 
repetitions $r$. Integration over $L$, Poisson's summation rule, and 
a stationary phase approximation, lead to the semiclassical Berry-Tabor 
formula \cite{berry_tabor}, 
expressing the density of states of a sphere in terms of classical 
periodic trajectories. Taking only into account
the smooth part of $d_L(E)$ in Eq.~(\ref{gamma_density}), 
we recover the well-known $1/R$ dependence \cite{yann_broglia} 
\begin{equation}\label{gamma_0}
\Gamma_0(R)=\frac{3\hbar}{4}\frac{v_{\text{F}}}{R} g(\xi)\, ,
\end{equation}
where $\xi=\hbar \omega_{\text{M}}/E_{\text{F}}$ , $g(\xi)$ is a smoothly 
decreasing function with $g(0)=1$, and  $v_{\text{F}}$ ($E_{\text{F}})$ is the 
Fermi velocity (energy).

Most interestingly, the oscillating part of the density of states gives 
rise to additional new features, namely an oscillation of $\Delta \Gamma(R)$ 
around $\Gamma_0(R)$ as a function of the radius. While the cross 
products between the smooth and the oscillating parts of $d_L(E)$ and 
$d_{L'}(E-\hbar\omega_{\text{M}})$ can be neglected after the energy 
integration, the product of the two oscillatory components, after doing the 
$L$ summation with the aid of Poisson's summation rule, yields
\begin{widetext}
\begin{eqnarray}
\Gamma_{\text{osc}}(R)&=&\frac{2\gamma}{\pi^2}\sum_{m=-\infty}^{\infty}
\int_{\rho_{\text{min}}}^{\rho_{\text{max}}}{\text{d}}\rho\, \rho 
\int_{-1/2}^{\rho}{\text{d}}y\,  \sum_{y'=y\pm 1} 
f(y')\sqrt{\rho^2-y^2}\sqrt{\rho'^2-y'^2} \nonumber \\ 
&& \times 
   \sum_{r=1}^\infty \sum_{r'=-\infty}^\infty
 (-1)^{r+r'} \cos\left[ r S(\rho,y)+ r' S(\rho',y')
   +2\pi m y\right]
\end{eqnarray}   
with $\rho'=\sqrt{\rho^2-(k_{\text{F}}R)^2\xi}$. 
The limits of the $\rho$-integral are given by 
$\rho_{\text{min}}=\max(k_{\text{F}}R,k_{\text{F}}R\xi)$ and 
$\rho_{\text{max}}=k_{\text{F}}R\sqrt{1+\xi}$. We have defined
$f(y')=y'+1$ for 
$y'=y-1$ and $f(y')=y'$ in the case $y'=y+1$ ($k_{\text{F}}$ is the Fermi
wave-vector). The dominant terms correspond 
to $m=0$, $r=-r'$ and $\bar{y}=\bar{y}'+1=\rho/(\rho-\rho')$, yielding    
\begin{equation}\label{gamma_osc}
\Gamma_{\text{osc}}(R)=\frac{2\gamma}{\pi^{3/2}}
\int_{\rho_{\text{min}}}^{\rho_{\text{max}}}\!\!{\text{d}}\rho\, 
     \frac{\rho^{7/2}\rho'^{3/2}\left((\rho-\rho')^2-1\right)^{5/4}}
           {(\rho-\rho')^4}  
\sum_{r=1}^{\infty} \frac{1}{\sqrt{r}}\, \cos\left[ 2r \left(
     \frac{(k_{\text{F}}R)^2\xi-2\bar{y}+1}
          {\sqrt{\rho^2-\bar{y}^2}+\sqrt{\rho'^2-\bar{y}'^2}}
          \right)\right] \, . 
\end{equation}
\end{widetext}   
Since $\rho$ is of the order of $k_{\text{F}}R$, and $k_{\text{F}}R\xi \gg 1$, 
the argument of the cosine is close to $2r k_{\text{F}}R \xi$, and 
(\ref{gamma_osc}) yields a contribution to $\Delta \Gamma$ which oscillates 
as a function of the radius of the nanoparticle.
It is important to note that this oscillatory behavior is due to the 
{\it angular-momentum restricted electron-hole density-density
correlation}, and therefore more subtle than the well-known oscillations
of the density of states with the energy.  

Fig.~\ref{fig:na} shows the result for 
$\Delta \Gamma=\Gamma_0+\Gamma_{\text{osc}}$ which is obtained by 
numerically evaluating the $\rho$-integral in Eq.~(\ref{gamma_osc}). 
The typical amplitude of the
oscillations can be estimated by the value 
$\Gamma_{\text{osc}}^{\text{max}}=6\sqrt{2\pi}\hbar^2/
(MR^2\sqrt{k_{\text{F}}R\xi^3})$, valid in the limit 
$k_{\text{F}}R \xi^2\ll 1$. 
This shows that the oscillating term represents an important correction to 
$\Gamma_0(R)$ for small particle sizes.

Higher repetitions $(r>1)$ of the periodic orbits are 
suppressed in Eq.~(\ref{gamma_osc}) because of the prefactor $1/\sqrt{r}$,
and the oscillations of the cosine as a function of $\rho$. 
Including a small intrinsic decoherence rate $\Gamma_{\text{i}}$ 
accounts for the coupling to the environment and leads to an additional 
factor $\exp(-\Gamma_{\text{i}} r R /h v_{\text{F}})$ \cite{lermesemi}, which
further reduces the oscillatory contributions from long trajectories 
with large repetition numbers $r$.  

%     
%%%%%%%%%%%%%%%%%%%% NUMERICS %%%%%%%%%%%%%%%%%%%%%%%%%%%%%%
%
\section{Numerical calculations}

In order to study the relevance of the oscillatory behavior in more general 
situations, we performed quantum numerical calculations within a spherically 
symmetric jellium model, using the linear response TDLDA \cite{ekardt,bertsch}.
The diagonalization of the Hamiltonian in the space of particle-hole 
excitations amounts to use the Random Phase Approximation in the 
self-consistent field with a local exchange potential. 
Further refinements of the numerical approach (like the Self-Interaction
Correction \cite{madjet}) do not provide substantial changes in the
effects we are interested in.
Because of the spherical symmetry of the model, we considered nanoparticle 
sizes corresponding to magic numbers of atoms where our numerical approach is 
expected to work the best \cite{nonspherical}. 

\subsection{Alkaline metals}

%
%%%%%%%%%%%%%%%%%%%% FIGURE SODIUM %%%%%%%%%%%%%%%%%%%%%%%%%
%       
\begin{figure}[bt]
\centerline{\epsfxsize=3in\epsffile{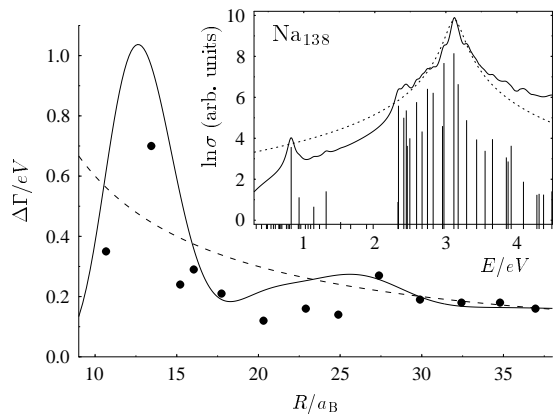}}
\caption{\label{fig:na} Linewidth $\Delta \Gamma(R)$, as a function of the 
radius (in units of the Bohr radius $a_{\text{B}}=0.53 \text{\AA}$), for Na 
nanoparticles containing between 20 and 832 atoms. The dots 
corresponding to TDLDA calculations are compared to the result of 
Eq.~(\ref{gamma_osc}) (full line). The smooth term $\Gamma_0(R)$ is 
given by Eq.~(\ref{gamma_0}) (dashed). Inset: Logarithm of the numerically 
calculated absorption cross-section for $\text{Na}_{138}$, showing a 
pronounced 
surface plasmon resonance, fitted by a Lorentzian (dotted line). The 
excited states are indicated by tick marks and their 
oscillation strengths are given by the height of the vertical lines.}
\end{figure}
%
%%%%%%%%%%%%%%%%%%%%% END FIGURE %%%%%%%%%%%%%%%%%%%%%%%%%%%
%       
A typical spectrum of the photo absorption cross-section 
(Eq.~(\ref{cross-section})) is shown in the inset of
Fig.~{\ref{fig:na}} for the example of a sodium cluster,
where a non-zero $\Gamma_{\text{i}}$ smears the singularities of the
spectrum. The collective excitation can be identified as the main peak
of the spectrum (note that the logarithm of $\sigma$ is plotted). 
The vertical lines are essentially given by {\it single}\/ electron-hole
excitations, except for the one at the center of the surface plasmon 
resonance, which is the {\it superposition}\/ of many such excitations. 
The surface plasmon linewidth $\Gamma$, represented by thick dots in 
Fig.~\ref{fig:na}, is obtained from a Lorentzian fit to the main peak. 
We chose an intrinsic width $\Gamma_{\text{i}}$ larger than the typical 
distance in the electron-hole spectrum and verified that 
$\Delta \Gamma(R)$ is approximately independent of the precise value 
of $\Gamma_{\text{i}}$. Confirming our analytical calculations, at 
small $R$ the numerical results show strong deviations from $\Gamma_0(R)$ 
(dotted line) and a non-monotonic behavior as a function of the radius.
Similar results are obtained for other alkaline metals. 
It should be mentioned that the phase of the semiclassical result from
Eq.~(\ref{gamma_osc}) is meaningful up to a constant shift which
was adjusted according to the numerical data.

\subsection{Noble metals and embedded clusters}

%
%%%%%%%%%%%%%%%%%%%% FIGURE SILVER %%%%%%%%%%%%%%%%%%%%%%%%%
%       
\begin{figure}[tb]
\centerline{\epsfxsize=3in\epsffile{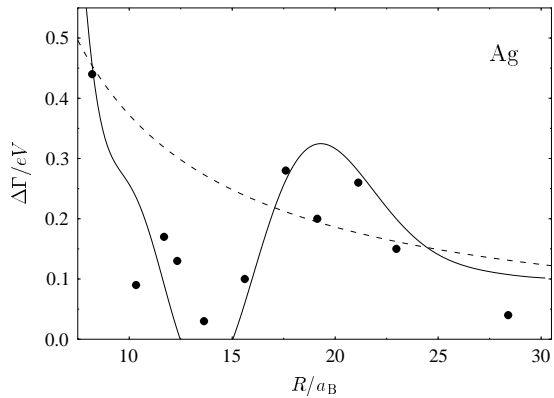}}
\caption{\label{fig:ag} The dots represent the numerically 
obtained surface plasmon linewidth for the case of silver nanoparticles 
embedded in an argon matrix, as a function of their radius $R$ in units of 
the Bohr radius. The dashed and the solid lines represent 
$\Gamma_0(R)$ and $\Delta \Gamma(R)$, respectively, reduced by a global 
factor of 3 with respect to Eqs.~(\ref{gamma_0}) and (\ref{gamma_osc}) 
(see text).}
\end{figure}
%
%%%%%%%%%%%%%%%%%%%%%% END FIGURE %%%%%%%%%%%%%%%%%%%%%%%%%%
%       
Noble metals are more difficult to describe, due to band-structure
effects and the fact that 
the comparison with experiments has to take into account the confinement 
of the particles in a matrix. In our numerical approach we included 
the contribution $\epsilon_{\text{d}}$ of the d-electrons to the dielectric 
function of the metal particles, as well as the dielectric constant 
$\epsilon_{\text{m}}$ of the matrix. For the resulting residual interaction 
we used a multipolar expansion as in Ref.~\cite{rubio}. We have checked that 
including an interstitial spacing (the so-called 3-$\epsilon$ model 
\cite{lerme}) does not change significantly our results. 

In Fig.~\ref{fig:ag} we show the numerical results (thick dots) for silver 
nanoparticles embedded in an inert matrix 
(Ar, $\epsilon_{\text{m}}\approx 1.75$), exhibiting pronounced oscillations of 
$\Delta \Gamma(R)$ with the particle size, much like in the case of 
alkaline metals. The smooth part of the linewidth $\Gamma_0(R)$ 
(Eq.~(\ref{gamma_0})), and our semiclassical result for 
$\Gamma_{\text{osc}}(R)$
(Eq.~(\ref{gamma_osc})) are an overall factor of 3 larger than
the numerical results. This discrepancy is not surprising given the
stringent approximations involved in both results, which are expected
to be less applicable for embedded noble metal clusters, than for free
alkaline metals. Nevertheless,
the analytically calculated $\Gamma_{\text{osc}}(R)$ exhibits 
oscillations with the correct period, and its relative size
(with respect to $\Gamma_0(R)$) agrees with the numerical results.

Reactive (non-inert) matrices like glass are often studied experimentally 
\cite{Kreibig}, but still more difficult to analyze. By taking into account 
a high-energy conduction band in the insulator \cite{persson}, we extended our 
numerical calculations to this case \cite{tbp}. These electronic states provide 
additional decay channels for the surface plasmon and its lifetime is 
considerably reduced. Our results for the Landau damping contribution to the 
linewidth of Ag nanoparticles embedded in SiO$_2$ amount to about one half 
of the experimental linewidth, but still show strong
size-dependent oscillations, as in the previous cases. 

%
% *********************** CONCLUSIONS ******************************
%

\section{Conclusions}

In conclusion, we have found a mechanism leading to an important
oscillatory contribution to the Landau damping linewidth of the collective 
surface plasmon excitation in small metallic nanoparticles. The
oscillations of the linewidth as a function of the particle size
arise from the oscillations of the angular-momentum restricted 
electron-hole density-density correlations, that we calculate
within a semiclassical approach. We obtain a quantitative agreement
between  analytical and numerical (TDLDA) results for free alkaline
metallic clusters, and a qualitative agreement for more complicated
cases.

A systematic experimental investigation of the line\-width
oscillations extending the results for specific situations
reported in Refs.~\cite{charle,brechignac,mochizuki} 
would provide an important step towards the understanding
of the rich physics involved in the collective excitations of finite systems. 
A small inhomogeneous broadening will not kill the effect if the size 
dispersion of the nanoparticles is smaller than the period of the 
oscillations (approximately $(k_{\text{F}} \xi)^{-1}$, according to 
Eq.~(\ref{gamma_osc})). Moreover, recent experimental developments
make it possible to study the homogeneous broadening of the surface 
plasmon also in electrically uncharged clusters \cite{fel,stietz}.
In the case of clusters embedded in a matrix, the oscillations of the
width of the plasmon can be seen not only as a function of the size but
also as a function of the dielectric constant of the environment. Changing
$\epsilon_{\text{m}}$ shifts the position of the plasmon, without considerably
modifying the single-particle part of the spectrum. Therefore the plasmon 
linewidth varies according to the position of the resonance with
respect to the particle-hole excitation spectrum.

%
% *********************** ACKNOWLEDGMENTS ******************************
%
\begin{acknowledgments}
We thank J.-Y.\ Bigot, M.\ Brack, J.\ Feldmann, L.\ Guidoni, H.\ Haberland, 
P.A.\ Hervieux, and D.\ Ullmo for helpful discussions. Financial support from 
the European Union through the TMR/RTN program is acknowledged.
\end{acknowledgments}
%
% *********************** REFERENCES ******************************
%


\begin{thebibliography}{99}
\bibitem{nonspherical} Incompletely filled shells result in a non-spherical 
  shape and a split resonance peak \cite{ellert}. This obvious shell effect 
  adds up to the oscillations that we have discussed.
\end{thebibliography}
\end{document}